\documentclass[conference,a4paper]{IEEEtran}

\usepackage{amsmath, amssymb}
\usepackage{graphicx}
\usepackage{url}

\usepackage{epsfig}
\usepackage{subcaption}
\usepackage{float}
\usepackage{caption}
\usepackage{booktabs}
\usepackage{array}
\usepackage{multirow}
\usepackage{tabularx}
\newcolumntype{Y}{>{\centering\arraybackslash}X}
\newcolumntype{R}{>{\raggedleft\arraybackslash}X}
\usepackage{arydshln}
\usepackage{tabu}
\usepackage{cancel}
\usepackage{algpseudocode}
\usepackage{amsmath}

\usepackage{xspace}
\usepackage{gensymb}
\usepackage{enumerate}
\usepackage{multicol}
\usepackage{enumitem}
\usepackage{balance}
\usepackage{siunitx}
\usepackage{cite}
\usepackage{glossaries}
\usepackage{caption}
\usepackage[table]{xcolor}
\usepackage{xcolor}

\definecolor{mitred}{HTML}{A31F34}  
\usepackage{bm}

\usepackage[
    colorlinks=true,  
    linkcolor=red,   
    urlcolor=pink,    
    citecolor=green    
]{hyperref}

\usepackage{adjustbox} 
\usepackage[linesnumbered,ruled,vlined]{algorithm2e}

\SetCommentSty{mycommfont}

\title{Accelerating Merge with Motion Vector Difference via Filter Difference Analysis for VVenC}

\author{double-blind submission}

\iftrue

\author{
\IEEEauthorblockN{
Xinmin Feng$^{1}$, Shengyang Xu$^{2}$, Jianhua Chen$^{2}$, Li Li$^{1}$, Dong Liu$^{1,\dagger}$, Feng Wu$^{1}$}
\IEEEauthorblockA{
$^{1}$University of Science and Technology of China,  $^{2}$Hupan Laboratory\\
xmfeng2000@mail.ustc.edu.cn; 
\{lil1, dongeliu, fengwu\}@ustc.edu.cn\\
[-2em]
\thanks{This work was supported by Hupan Laboratory through Hupan Laboratory Research Intern Program.}
}
}

\fi
\begin{document}
\sloppy

\maketitle
\begin{abstract}
Merge with Motion Vector Difference (MMVD) is a key coding tool in Versatile Video Coding for improving motion prediction accuracy. However, its exhaustive search strategy imposes a significant computational burden on the encoder. To address this issue, we propose a novel fast MMVD algorithm for the VVenC encoder based on fractional motion vector filter difference analysis. By approximating the 8-tap interpolation filter with a 2-tap filter, we derive a criterion based on spatial gradients and prediction residuals for estimating the potential gain of MMVD candidates. We further generalize this criterion to accommodate both shifted integer reference samples and 2D separable filtering. To minimize the overhead of the proposed method, we introduce implementation optimizations, including symmetric offset inference and cross-shaped downsampled dot-product computation. Compared with existing fast MMVD algorithms in VVenC, our method reduces the average MMVD search ratio from 21.07\% to 11.05\% and decreases the efficiency-complexity metric $\eta$ from 11.79 to 7.10 under the fast preset.

\end{abstract}

\begin{IEEEkeywords}
Versatile Video Coding, Merge with Motion Vector Difference, Fast algorithm.
\end{IEEEkeywords}

\section{Introduction}
\label{sec:introduction}

Merge mode is a key technique in modern video coding standards for exploiting spatial and temporal redundancies. Adopted in both High Efficiency Video Coding (HEVC)~\cite{sullivan2012overview} and Versatile Video Coding (VVC)~\cite{bross2021overview}, it reduces motion vector (MV) bitrate by reusing motion information from a list of merge candidates for the current coding unit (CU).
In HEVC, this list is built from spatial and temporal neighbors. VVC enhances it with history-based~\cite{li2019history} and  subblock-based motion vector prediction~\cite{liao2025subblock}, geometric partitioning~\cite{gao2020geometric}, and combined inter-intra prediction~\cite{Pham2019CE10}. It also incorporates decoder-side motion vector refinement~\cite{gao2020decoder} and Merge with Motion Vector Difference (MMVD)~\cite{jeong2020merge} for higher motion precision.

Unlike standard merge mode, MMVD allows signaling an additional motion vector difference (MVD) to refine the selected base motion. Specifically, MMVD selects one of the first two merge candidates as the base motion. The refinement is represented by a direction and a distance. Four directions, including 0°, 90°, 180°, and 270°, are supported, with the direction index signaled in the bitstream. As shown in Table \ref{tab:steps}, two predefined distance tables~\cite{Liu2019AHG11}, each containing eight entries, define the offset magnitude. The encoder selects one table at the picture level and signals a distance index for the MVD. Though effective in coding, MMVD significantly increases encoding complexity because of exhaustive search over all direction–distance combinations. Specifically, enabling MMVD in VVenC introduces encoding time overheads of \textbf{36.39\%} and \textbf{21.20\%} under the faster and fast presets, respectively, hindering its usage in low-latency scenarios.

\begin{table}[t!]
    \caption{Distance tables (in unit of luma samples) used in MMVD.}
    \centering
        \begin{adjustbox}{width=0.3\textwidth, totalheight=\textheight, keepaspectratio}
    \begin{tabular}{lccccccccc}
    \toprule
    \textbf{Index} & 0 & 1 & 2 & 3 & 4 & 5 & 6 & 7  \\
    \midrule
    \textbf{Table 1}& 1/4 & 1/2 & 1 & 2 & 4 & 8 & 16 & 32\\
    \textbf{Table 2}& 1 & 2 & 4 & 8 & 16 & 32 & 64 & 128\\

    \bottomrule
    \end{tabular}
    \end{adjustbox}
    \label{tab:steps}
\end{table}

\begin{figure}[t]
    \centering
    \begin{subfigure}[!t]{0.1\textwidth}
        \centering
        \includegraphics[width=\linewidth]{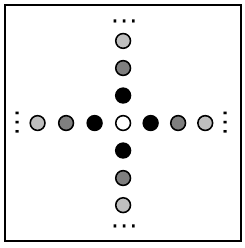}
        \caption{}
        \label{fig:mmvd_search}
    \end{subfigure}
    \hspace{3pt}
    \begin{subfigure}[!t]{0.22\textwidth}
        \centering
        \includegraphics[width=\linewidth]{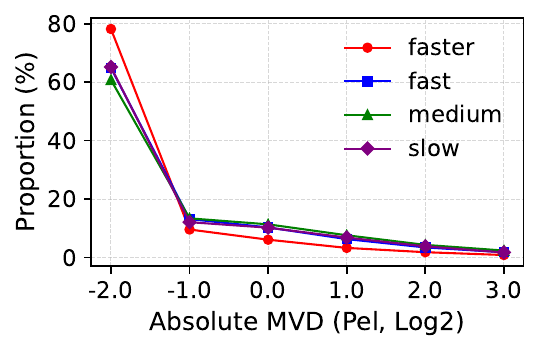}
        \caption{}
        \label{fig:second_sub}
    \end{subfigure}
    \caption{Illustration of MMVD search process. (a) MMVD Search Process. (b) Distribution of MMVD steps.}
    \vspace{-10pt}
    \label{fig:mmvd_fig1}
\end{figure}

In this paper, we focus on accelerating MMVD from the perspective of fractional motion vector filter difference analysis. As shown in Fig. \ref{fig:mmvd_fig1}(b), existing MMVD modes still require evaluating a large portion of quarter-pixel MVD candidates, which accounts for 60\%$\sim$80\% of total trials. To alleviate this issue, we propose a residual-gradient criterion that estimates whether a given MMVD direction is likely to provide coding gain. The criterion is first derived for a 1D interpolation case where regular merge and MMVD use the same integer-positioned reference samples. We then extend it to practical cases where MMVD offsets crosses an integer-sample boundary, causing the interpolation support to shift by one sample, as well as to 2D separable filtering. In addition, we introduce several engineering optimizations, including symmetric offset inference and efficient dot-product computation, to reduce the overhead of the proposed method itself. Experimental results on the VVenC encoder demonstrate that, across multiple presets, our approach achieves a better trade-off between encoding complexity and RD performance than existing MMVD acceleration schemes.

\section{Methodology}
\label{sec:method}

In this section, we present an efficient method for accelerating MMVD in VVC by reusing the regular-merge prediction and the spatial gradients of the reference blocks. The core idea is to approximate the 8-tap fractional motion compensation filters with 2-tap filters and derive a lightweight criterion for early termination of non-promising MMVD directions.

\subsection{MMVD Prediction Approximation}
\label{ssec:approximation}

\begin{figure}[t]
    \centering
    \includegraphics[width=0.7\linewidth]{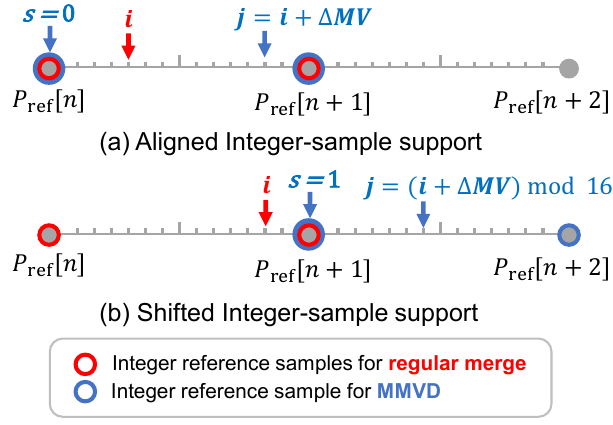}
    \caption{{Illustration of the integer-positioned reference samples used by the regular-merge and MMVD predictions using a 2-tap interpolation filter as an example. The fractional MV of regular merge and MMVD, denoted as $i$ and $j$, respectively, may use either (a) the same pair of integer-positioned reference samples, referred to as the \textit{aligned-support} case, or (b) different sample pairs after a one-sample support shift, referred to as the \textit{shifted-support} case.}
    }
    \vspace{-10pt}
    \label{fig:support}
\end{figure}

{We first distinguish two cases based on whether MMVD changes the integer-positioned reference samples used for interpolation, as shown in Fig.~\ref{fig:support}. Specifically, let $i\in\{0,\ldots,15\}$ denote the fractional phase of the regular-merge MV along one dimension, and let $\Delta$ denote the MMVD offset in $1/16$-sample units. The refined position of MMVD $q:=i+\Delta$ can be decomposed as
$s=\left\lfloor q/16\right\rfloor$ and $j=q-16s\in\{0,\ldots,15\}$,
where $j$ is the refined fractional phase and $s$ is the corresponding integer-position shift.  $s=0$ indicates that regular merge and MMVD use the same pair of integer-positioned reference samples, whereas $s=\pm1$ indicates that MMVD uses an adjacent pair. In this paper, we refer to these two cases as the \textit{aligned-support} and \textit{shifted-support} cases, respectively.}

\begin{figure}[t]
    \centering
    \includegraphics[width=\linewidth]{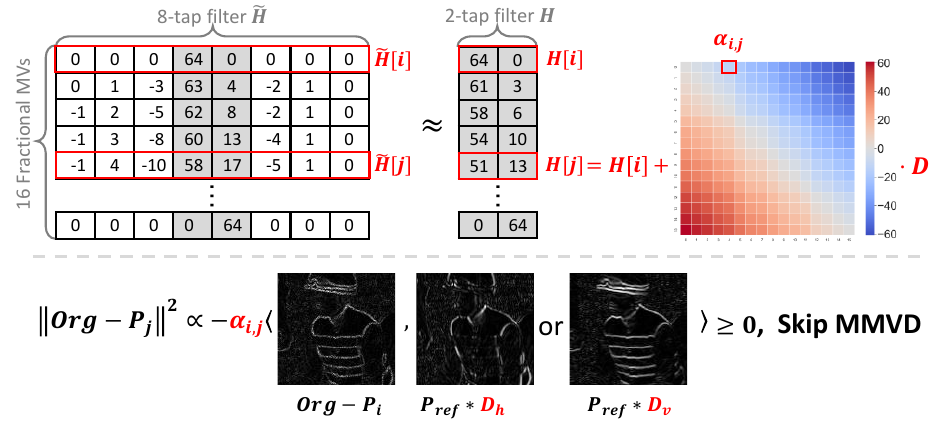}
    \caption{Illustration of the proposed fast MMVD algorithm for the aligned-support case. Given the original block $\bm O$, regular-merge phase $i$, and prediction $\bm P_i=\bm P_{\text{ref}}*\bm H[i]$, the MMVD prediction at phase $j$ is approximated as $\bm P_j\approx\bm P_i+\alpha_{i,j}(\bm P_{\text{ref}}*\bm D)$. The candidate is skipped if $-\alpha_{i,j}\langle\bm O-\bm P_i,\bm P_{\text{ref}}*\bm D\rangle\ge0$. }
    \label{fig:framework}
\end{figure}

{In the \textbf{\textit{aligned-support}} case}, both predictions are interpolated from the same reference block:
$\bm P_i=\bm P_{\text{ref}}*\widetilde{\bm H}[i]$ and
$\bm P_j=\bm P_{\text{ref}}*\widetilde{\bm H}[j]$,
where $\widetilde{\bm H}[i]$ and $\widetilde{\bm H}[j]$ are the corresponding 8-tap DCT-IF filters. 
As shown in Fig.~\ref{fig:framework}, the main coefficients of the interpolation filter are concentrated on two neighboring integer samples. Therefore, we approximate the 8-tap filter $\widetilde{\bm H}$ with a 2-tap filter
$\bm H=[H^{(0)},H^{(1)}]$, where $H^{(0)}+H^{(1)}=64$.
The difference between the two approximated filters is
\begin{equation}
\Delta\bm H_{i,j}
=\bm H[j]-\bm H[i]
=\alpha_{i,j}\bm D,\quad
\bm D=[-1,1],
\label{eq}
\end{equation}
where $\alpha_{i,j}=H_i^{(0)}-H_j^{(0)}$ is precomputed for each phase pair. Thus, the MMVD prediction can be approximated as
\begin{equation}
\bm P_j
\approx\bm P_i+\alpha_{i,j}(\bm P_{\text{ref}}*\bm D)
=\bm P_i+\alpha_{i,j}\bm g_{\text{ref}},
\label{eq}
\end{equation}
where $\bm g_{\text{ref}}$ is the spatial gradient of the reference block along the MMVD direction. Let $\bm O$ be the original block and $\bm e_i=\bm O-\bm P_i$ be the residual of the regular-merge candidate. The distortion of the MMVD candidate is approximated as
\begin{equation}
d_{i,j}
= \|\bm{O}-\bm{P}_j\|^2
\approx \|\bm{e}_i-\alpha_{i,j}\bm{g}_\text{ref}\|^2.
\label{eq}
\end{equation}
Expanding this term yields
\begin{equation}
d_{i,j}
\approx
\|\bm{e}_i\|^2
+ \alpha_{i,j}^2\|\bm{g}_\text{ref}\|^2
- 2\alpha_{i,j}\langle \bm{e}_i,\bm{g}_\text{ref}\rangle.
\label{eq:first_order2}
\end{equation}
The key observation is that the potential gain of the MMVD candidate is mainly determined by the sign of the inner-product term $-\langle \bm{e}_i, \bm{g}_\text{ref} \rangle$. If $-\alpha_{i,j} \langle \bm{e}_i, \bm{g}_\text{ref} \rangle \geq 0$, the MMVD candidate is unlikely to achieve lower distortion than regular merge, and its evaluation can therefore be skipped.

{We further extend the analysis to the \textbf{\textit{shifted-support}} case.} As shown in Fig. \ref{fig:support}(b), the union of the two supports contains three integer samples, which can be factorized using the same differential kernel:
\begin{equation}
\bm{\Delta H}_{i,j}
=
\begin{cases}
-\bm{D} * [{H}_j^{(0)},\,{H}_i^{(1)}],
\text{ leftward shift},\\
\phantom{-}\bm{D} * [{H}_i^{(0)},\,{H}_j^{(1)}],
\text{ rightward shift}.
\end{cases}
\end{equation}
Thus, the gradient $\bm g_\text{ref}$ only needs to be computed once. The final prediction adjustment can then be obtained by applying a simple 2-tap filter to $\bm{g}_{\text{ref}}$, avoiding direct 3-tap filtering on the original reference block. Following the same first-order criterion in Eq. (\ref{eq:first_order2}), the candidate is skipped if the corresponding residual-gradient response is positive: $-\left<\bm{e}_i, \bm{P}_{\text{ref}} * \Delta\bm{H}_{i,j}\right> \ge 0$.
Notably, $\bm{P}_{\text{ref}}$ should also be shifted by one pixel according to $s$ before computing $\bm{g}_{\text{ref}}$ to align with the new integer position of the MMVD candidate.

\begin{algorithm}[h]
\small
\DontPrintSemicolon 
\caption{Proposed Fast MMVD Algorithm}
\label{alg:algorithm-pipeline}
\KwIn{Source block $\bm{O}$, biprediction reference blocks $\{\bm{P}_{\text{ref}}^{(0)}, \bm{P}_{\text{ref}}^{(1)}\}$, BCW weights $\{w_0, w_1\}$, reconstruction of regular merge mode $\bm{P}_{\text{merge}}$, motion vector $(i_h, i_v)$, MMVD offset direction $(\Delta_h, \Delta_v) \in \{(\Delta d, 0), (-\Delta d, 0), (0, \Delta d), (0, -\Delta d)\}$}
\KwOut{Whether to early-skip the current MMVD direction}

Initialize $\textit{dot} \gets 0$\;
\For{$i \in \{0, 1\}$}{
    $\bm{e} \gets \bm{O} - \bm{P}_{\text{merge}}$, \quad $\bm{g}_h^{(i)} \gets \bm{D}_h * \bm{P}_{\text{ref}}^{(i)}$\;
    $s \gets \left\lfloor {(i_h + \Delta_h)}/{16} \right\rfloor$\;
    \eIf{$s = 0$}{
        $\bm{D}_f \gets H_{i_h+\Delta_h}^{(0)} - H_{i_h}^{(0)}$\tcp*[r]{aligned-support} 
    }{\eIf{$s = 1$}{
        $\bm{D}_f \gets  -[H_{i_h+\Delta_h}^{(0)},\, H_{i_h}^{(0)}] $\tcp*[r]{leftward shifted-support}
    }{
        $\bm{D}_f \gets [H_{i_h}^{(0)},\,H_{i_h+\Delta_h}^{(0)}]$\tcp*[r]{rightward shifted-support}
    }}
    $\textit{dot} \gets \textit{dot} - w_i \langle \bm{e}, \bm{D}_f * \bm{g}_h^{(i)} \rangle$\;
}
\KwRet{$\textit{dot} \ge 0$}\tcp*[r]{Skip this direction} 
\vspace{0.5em}
\footnotesize\textit{Note: For brevity, we only take bidirectional references and horizontal MMVD offsets as example.}
\end{algorithm}

\subsection{Extension to 2D Separable Filtering}

We next extend the 1D analysis to the 2D separable interpolation
used in VVC, where the regular-merge with the fractional MV $(i_h, i_v)$ and MMVD predictions with the fractional MV $(j_h, j_v)$ are
\begin{align}
    \bm P_i
    &=\bm H_v[i_v]*(\bm H_h[i_h]*\bm P_{\text{ref}}),\\
    \bm P_j
    &=\bm H_v[j_v]*(\bm H_h[j_h]*\bm P_{\text{ref}}).
\end{align}
Using the 2-tap approximation,
\begin{align}
    \bm{H}_h[j_h] \approx \bm{H}_h[i_h] + \alpha_h\bm{D}_h, 
    \bm{H}_v[j_v] \approx \bm{H}_v[i_v] + \alpha_v\bm{D}_v,
\end{align}
where $\alpha_h$ and $\alpha_v$ are pre-computed horizontal and vertical scalars.
Substituting these approximations into $\bm{P}_j$ gives
\begin{equation}
\begin{aligned}
\bm{P}_j
&\approx
(\bm{H}_v[i_v]+\alpha_v\bm{D}_v)
*
\left((\bm{H}_h[i_h]+\alpha_h\bm{D}_h)*\bm{P}_{\text{ref}}\right) \\
&= \bm{P}_i
+ \alpha_h\bm{H}_v[i_v]*(\bm{D}_h*\bm{P}_{\text{ref}}) \\
&\quad
+ \alpha_v\bm{D}_v*(\bm{H}_h[i_h]*\bm{P}_{\text{ref}})
+ \cancel{\alpha_h\alpha_v\bm{D}_v*(\bm{D}_h*\bm{P}_{\text{ref}})} .
\end{aligned}
\end{equation}
Since standard MMVD only searches four cardinal directions, only one of $\alpha_h$ and $\alpha_v$ is non-zero for each candidate. Thus, the cross-term vanishes. For a horizontal MMVD candidate,
\begin{equation}
    \bm{P}_j
    \approx
    \bm{P}_i
    + \alpha_h\bm{H}_v[i_v]*(\bm{D}_h*\bm{P}_{\text{ref}}).
\end{equation}
For a vertical MMVD candidate,
\begin{equation}
    \bm{P}_j
    \approx
    \bm{P}_i
    + \alpha_v\bm{D}_v*(\bm{H}_h[i_h]*\bm{P}_{\text{ref}}).
\end{equation}
Hence, the residual-gradient criterion derived in Eq.~\eqref{eq:first_order2} can also be applied to both horizontal and vertical MMVD directions.
The complete implementation is summarized in Algorithm~\ref{alg:algorithm-pipeline}, where we use $s = \left\lfloor {(i_h+\Delta_h)}/{16} \right\rfloor$ to distinguish the aligned- and shifted-support cases. Different cases are finally converted into the same dot-product form between the residual of regular merge modes and the spatial gradient of reference blocks, as shown in Line~12 of Algorithm~\ref{alg:algorithm-pipeline}.

\subsection{Implementation Optimizations}\label{ssec:engineering}
To reduce the complexity overhead of the proposed algorithm, we introduce two implementation optimizations.
\textbf{(1) Symmetric Offset Inference:} For a pair of symmetric MMVD offsets with opposite signs along the same axis, e.g., $(\Delta d, 0)$ and $(-\Delta d, 0)$, when one offset leads to the shifted-support case while its counterpart remains in the aligned-support case, we explicitly compute the residual-gradient response for the aligned-support offset and infer the response of the shifted-support offset by sign inversion. The inferred response is then used to decide whether the corresponding MMVD candidate can be skipped.
\textbf{(2) Efficient Dot Product Calculation:} Instead of computing the dot product over the entire block, we restrict the calculation to a central cross-shaped region. Specifically, the middle row is used for horizontal MMVD offsets, and the middle column is used for vertical ones.

\begin{algorithm}[t]
\small
\DontPrintSemicolon
\caption{Integration of the Proposed Algorithm}
\label{alg:fast_mmvd_search}
\KwIn{Merge candidate list, cost of last entry $C_{\text{last}}$}
\KwOut{Optimal MMVD vector and its cost}

Initialize bestDir $\gets \varnothing$\;
\For{step $\in \{0, 1, \dots, 7\}$}{
    minCost $\gets \infty$\;
    \For{dir $\in \{\text{right}, \text{left}, \text{down}, \text{up}\}$}{
        Determine the MMVD offset $(\Delta_h, \Delta_v)$ from $(step, dir)$\;
        \textcolor{mitred}{
        \If{Algorithm~\ref{alg:algorithm-pipeline} returns true for $(\Delta_h, \Delta_v)$}{
            Continue\tcp*[r]{\textbf{\textit{Proposed Algorithm}}}
        }}
        \If{step $\geq 1$ \textbf{and} dir $\neq$ bestDir}{  
            Continue\tcp*[r]{(\textit{{\textbf{Fast2}}})}
        }
        Compute cost $c$ for (step, dir)\;
        \If{$c <$ minCost}{
            minCost $\gets c$\;
            \If{step $= 0$}{
                bestDir $\gets$ dir\;
            }
        }
    }
    \If{minCost $> C_{\text{last}}$}{
        Break\tcp*[r]{(\textit{{\textbf{Fast3}}})}
    }
}
\KwRet{best MMVD result (bestDir, step, minCost)}\;
\end{algorithm}

\section{Experiments}
\label{sec:experiments}
\subsection{Experimental Setup}
We evaluate the RD performance and encoding complexity of enabling MMVD using the VVenC-1.14.0 encoder across four presets: faster, fast, medium, and slow. The proposed method is implemented on top of the existing MMVD fast-search strategies in VVenC, as shown in Algorithm~\ref{alg:fast_mmvd_search}. The original fast strategies include:
\begin{itemize}
\item (\textit{\textbf{Fast1}}) MMVD is evaluated only when the cost of its base merge candidate ranks within the top two positions of the coarse search list.
\item (\textit{\textbf{Fast2}}) After the first step size is tested, only the best direction is retained for subsequent MMVD evaluations.
\item (\textit{\textbf{Fast3}}) Further MMVD evaluations are skipped when the best MMVD cost is higher than the cost of the last candidate in the coarse search list.
\end{itemize}
Here, we compare three MMVD-enabled configurations. Specifically, \textbf{\textit{Baseline}} refers to full MMVD search without fast pruning, {\textbf{\textit{+Fast123}} refers to MMVD accelerated by the above three strategies}, and \textbf{\textit{+Ours}} further integrates the proposed algorithm on top of \textbf{\textit{+Fast123}}.
Experiments are conducted under Random Access configurations following the VVC common test conditions~\cite{bossen2020jvet}, using four quantization parameters: 22, 27, 32, and 37. The test sequences cover JVET CTC Classes A1, A2, B, C, and D. Compression performance is evaluated using the BD-Rate~\cite{bjontegaard2001calculation}. Encoding complexity is calculated as:
\begin{equation}
\Delta T_\text{enc} = \frac{T_\text{test} - T_\text{anchor}}{T_\text{anchor}} \times 100\%,
\end{equation}
where $T_\text{test}$ denotes the encoding time of the tested MMVD-enabled configuration, and $T_\text{anchor}$ denotes the encoding time of the original VVenC-1.14.0 encoder with MMVD disabled. For PSNR and SSIM~\cite{wang2004image}, we report the weighted Bjøntegaard Delta rate using Y:Cb weights of 6:1:1. For VMAF~\cite{li2016toward}, BD-rate is computed using the luma-domain VMAF score.
To assess the cost-effectiveness of enabling MMVD, we further compute the ratio between encoding-time overhead and the average BD-rate gain over the reported quality metrics: $\eta = -{\Delta T_\text{enc}}/{\text{BD-rate}},$ where a lower $\eta$ indicates that less encoding-time overhead is required to obtain each unit of BD-rate gain, corresponding to a better trade-off between encoding complexity and RD performance.

\begin{table}[!t]
\centering
\caption{Experimental results for different MMVD fast algorithms across different presets. \textcolor{red}{\textbf{Bold}}: best performance. \textcolor{blue}{\underline{Underline}}: Second Best Performance.}
\label{tab:results_long}
\small
\begin{adjustbox}{width=0.45\textwidth, totalheight=\textheight, keepaspectratio}
\begin{tabular}{l l l c c c c c}
\toprule\toprule
 & \multicolumn{2}{c}{\multirow{2}{*}{\textbf{Method}}} & \multicolumn{3}{c}{\textbf{BD-Rate} (\%)} & \multirow{2}{*}{$\Delta \bm{T}_{\text{enc}}$ (\%)} & \multirow{2}{*}{$\bm\eta\ (\downarrow)$} \\
 & & & \textbf{SSIM} & \textbf{PSNR} & \textbf{VMAF} & & \\
\midrule\midrule

\multirow{8}{*}{\rotatebox{90}{\textbf{\textit{Faster Preset}}}}
& \textit{\textbf{Baseline}}      & Average & -1.43 & -1.12 & -0.12 & 36.39 & 40.89 \\
& \textit{\textbf{+Fast123}}   & Average & -0.87 & -0.79 & -0.18 & 7.20  & \textcolor{blue}{\underline{11.74}} \\
\cmidrule(lr){2-8}

& \multirow{6}{*}{\textbf{\textit{+Ours}}} 
& \textcolor{gray}{A1} & \textcolor{gray}{-0.80} & \textcolor{gray}{-0.52} & \textcolor{gray}{-0.38} & \textcolor{gray}{4.17} & \textcolor{gray}{7.36} \\
&  & \textcolor{gray}{A2} & \textcolor{gray}{-0.76} & \textcolor{gray}{-0.69} & \textcolor{gray}{-0.08} & \textcolor{gray}{4.72} & \textcolor{gray}{9.25} \\
&  & \textcolor{gray}B  & \textcolor{gray}{-0.62} & \textcolor{gray}{-0.78} & \textcolor{gray}{-0.20} & \textcolor{gray}{4.64} & \textcolor{gray}{8.70} \\
&  & \textcolor{gray}C  & \textcolor{gray}{-0.42} & \textcolor{gray}{-0.53} & \textcolor{gray}{-0.11} & \textcolor{gray}{4.98} & \textcolor{gray}{14.09} \\
&  & \textcolor{gray}D  & \textcolor{gray}{-0.86} & \textcolor{gray}{-1.09} & \textcolor{gray}{-0.58} & \textcolor{gray}{4.68} & \textcolor{gray}{5.55} \\
&   & Average & -0.69 & -0.72 & -0.27 & 4.64 & \textcolor{red}{\textbf{8.26}} \\
\midrule\midrule

\multirow{8}{*}{\rotatebox{90}{\textbf{\textit{Fast Preset}}}}
& \textbf{\textit{Baseline}}      & Average & -0.90 & -0.64 & -0.26 & 21.10 & 35.17 \\
& \textit{\textbf{+Fast123}}   & Average & -0.63 & -0.41 & -0.11 & 4.52  & \textcolor{blue}{\underline{11.79}} \\
\cmidrule(lr){2-8}

& \multirow{6}{*}{\textbf{\textit{+Ours}}} 
& \textcolor{gray}{A1} & \textcolor{gray}{-0.78} & \textcolor{gray}{-0.54} & \textcolor{gray}{-0.59} & \textcolor{gray}{2.91} & \textcolor{gray}{4.57} \\
&  & \textcolor{gray}{A2} & \textcolor{gray}{-0.59} & \textcolor{gray}{-0.35} & \textcolor{gray}{-0.10} & \textcolor{gray}{1.69} & \textcolor{gray}{4.88} \\
&  & \textcolor{gray}B  & \textcolor{gray}{-0.54} & \textcolor{gray}{-0.41} & \textcolor{gray}{-0.29} & \textcolor{gray}{2.87} & \textcolor{gray}{6.94} \\
&  & \textcolor{gray}C  & \textcolor{gray}{-0.19} & \textcolor{gray}{-0.11} & \textcolor{gray}{0.18}  & \textcolor{gray}{2.49} & \textcolor{gray}{62.25} \\
&  & \textcolor{gray}D  & \textcolor{gray}{-0.35} & \textcolor{gray}{-0.32} & \textcolor{gray}{0.27}  & \textcolor{gray}{1.18} & \textcolor{gray}{8.85} \\
&   & Average & -0.49 & -0.35 & -0.11 & 2.23 & \textcolor{red}{\textbf{7.10}} \\
\midrule\midrule

\multirow{8}{*}{\rotatebox{90}{\textbf{\textit{Medium Preset}}}}
& \textbf{\textit{Baseline}}      & Average & -0.55 & -0.30 & 0.07 & 16.59 & 63.81 \\
& \textit{\textbf{+Fast123}}    & Average & -0.39 & -0.19 & 0.05 & 3.19  & \textcolor{blue}{\underline{18.06}} \\
\cmidrule(lr){2-8}

& \multirow{6}{*}{\textbf{\textit{+Ours}}} 
& \textcolor{gray}{A1} & \textcolor{gray}{-0.19} & \textcolor{gray}{-0.07} & \textcolor{gray}{-0.23} & \textcolor{gray}{1.94} & \textcolor{gray}{11.88} \\
&  & \textcolor{gray}{A2} & \textcolor{gray}{-0.26} & \textcolor{gray}{-0.11} & \textcolor{gray}{-0.03} & \textcolor{gray}{1.30} & \textcolor{gray}{9.75} \\
&  & \textcolor{gray}B  & \textcolor{gray}{-0.41} & \textcolor{gray}{-0.27} & \textcolor{gray}{-0.14} & \textcolor{gray}{2.10} & \textcolor{gray}{7.68} \\
&  & \textcolor{gray}C  & \textcolor{gray}{-0.17} & \textcolor{gray}{-0.10} & \textcolor{gray}{0.25}  & \textcolor{gray}{1.90} & \textcolor{gray}{285.00} \\
&  & \textcolor{gray}D  & \textcolor{gray}{-0.26} & \textcolor{gray}{-0.18} & \textcolor{gray}{0.51}  & \textcolor{gray}{1.64} & \textcolor{gray}{-70.29} \\
&   & Average & -0.26 & -0.15 & 0.07 & 1.78 & \textcolor{red}{\textbf{16.05}} \\
\midrule\midrule

\multirow{8}{*}{\rotatebox{90}{\textit{\textbf{Slow Preset}}}}
& \textbf{\textit{Baseline}}      & Average & -0.48 & -0.39 & -0.24 & 5.21 & 14.08 \\
& \textit{\textbf{+Fast123}}    & Average & -0.36 & -0.29 & -0.16 & 1.15 & \textcolor{blue}{\underline{4.26}} \\
\cmidrule(lr){2-8}

& \multirow{6}{*}{\textbf{\textit{+Ours}}} 
& \textcolor{gray}{A1} & \textcolor{gray}{-0.28} & \textcolor{gray}{-0.19} & \textcolor{gray}{-0.16} & \textcolor{gray}{0.76} & \textcolor{gray}{3.62} \\
&  & \textcolor{gray}{A2} & \textcolor{gray}{-0.27} & \textcolor{gray}{-0.24} & \textcolor{gray}{0.07}  & \textcolor{gray}{0.74} & \textcolor{gray}{5.05} \\
&  & \textcolor{gray}B  & \textcolor{gray}{-0.32} & \textcolor{gray}{-0.24} & \textcolor{gray}{-0.21} & \textcolor{gray}{0.64} & \textcolor{gray}{2.49} \\
&  & \textcolor{gray}C  & \textcolor{gray}{-0.10} & \textcolor{gray}{-0.08} & \textcolor{gray}{-0.15} & \textcolor{gray}{0.71} & \textcolor{gray}{6.45} \\
&  & \textcolor{gray}D  & \textcolor{gray}{-0.36} & \textcolor{gray}{-0.31} & \textcolor{gray}{-0.12} & \textcolor{gray}{0.10} & \textcolor{gray}{0.38} \\
&   & Average & -0.27 & -0.21 & -0.11 & 0.59 &\textcolor{red}{\textbf{2.99}}  \\

\bottomrule\bottomrule
\end{tabular}
\end{adjustbox}
\end{table}

\subsection{Experimental Results}
Table~\ref{tab:results_long} summarizes the RD performance and encoding complexity of the three MMVD-enabled configurations. Full MMVD search provides the largest coding gain but introduces substantial encoding overhead, especially for the faster, fast, and medium presets. The existing VVenC fast-search strategies greatly reduce this overhead, while the proposed method further improves the efficiency--complexity trade-off across all presets. Compared with \textbf{\textit{+Fast123}}, \textbf{\textit{+Ours}} reduces the encoding-time overhead from 7.20\% to 4.64\% under the faster preset and from 4.52\% to 2.23\% under the fast preset. Meanwhile, the corresponding $\eta$ values decrease from 11.74 to 8.26 and from 11.79 to 7.10, respectively. Similar trends are observed for the medium and slow presets, where the proposed method achieves the lowest $\eta$. These results indicate that the proposed acceleration strategy effectively removes less promising MMVD candidates while largely preserving the coding benefit of MMVD.

\begin{table}[!t]
    \centering
    \caption{Ablation study on the implementation optimizations.}
    \label{tab:ablation_study}
    \small 
    \begin{adjustbox}{width=0.45\textwidth, totalheight=\textheight, keepaspectratio}
    \begin{tabular}{lcccccccccccc}
        \toprule
         \multirow{2}{*}{\textbf{SIMD}} & \multirow{2}{*}{\textbf{SymOffset}} & \multirow{2}{*}{\textbf{CrossDot}}  & \multicolumn{3}{c}{\textbf{BD-Rate} (\%)} & \multirow{2}{*}{$\Delta \bm{T}_{\text{enc}}$ (\%)} & \multirow{2}{*}{$\bm\eta\ (\downarrow)$} \\
         & & & \textbf{SSIM} & \textbf{PSNR} & \textbf{VMAF} & & \\        
        \midrule
         & & & -0.54 & -0.37 & -0.10 & 4.21 &  12.50 \\
        \checkmark& & & -0.54 & -0.37 & -0.10 & 3.26 &  9.68 \\
         \checkmark&\checkmark& & -0.51 & -0.36 & -0.07 & 2.93 &  9.35  \\
        \checkmark& \checkmark&\checkmark & -0.49 & -0.35 & -0.11 & 2.23 & 7.10  \\
        \bottomrule
    \end{tabular}
    \end{adjustbox}
\end{table}

To evaluate the implementation optimizations introduced in Section~\ref{ssec:engineering}, we conduct an ablation study using the fast preset. Starting from the block-level dot-product implementation, we sequentially apply three techniques: Single Instruction Multiple Data (SIMD) optimization, symmetric offset inference for reducing redundant computations of shifted-support cases, and cross-shaped dot-product calculation. As shown in Table~\ref{tab:ablation_study}, these optimizations have negligible impact on coding performance while significantly reducing computational overhead, lowering the encoding-time increase from 4.21\% to 2.23\%, validating the proposed implementation optimizations.

Moreover, Table~\ref{tab:mmvd_attempts} reports the average number of MMVD search attempts for each CTU across four presets. Relative to full MMVD search, the existing fast strategies reduce the attempts to 21.07\%, and the proposed method further lowers this ratio to 11.05\%. Thus, only about half of the candidates retained by \textbf{\textit{+Fast123}} are finally evaluated, i.e., $r = {11.05}/{21.07} \approx 0.52$.
For a coding block of size $W \times H$, the original MMVD evaluation requires approximately $2NWHK$ operations, where $N$ is the number of MMVD candidates and $K$ is the interpolation filter length. With pruning, only an $r$ portion of candidates requires full evaluation, while the algorithm adds $4(W+H)$ operations for cross-shaped 2-tap filtering and dot-product computation. Thus, the overall complexity is approximated as $2rNWHK + 4(W+H)$.
Since this additional $\mathcal{O}(W+H)$ overhead is much smaller than full block-wise motion compensation for typical CU sizes, the proposed method substantially reduces MMVD search complexity after \textbf{\textit{+Fast123}}.
\begin{table}[!t]
    \centering
    \caption{Average number of MMVD search attempts per CTU.}
    \label{tab:mmvd_attempts}
    \small 
    \begin{adjustbox}{width=0.48\textwidth, totalheight=\textheight, keepaspectratio}
        \begin{tabular}{lccccc}
            \toprule
            \textbf{Method} & \textbf{Faster} & \textbf{Fast} & \textbf{Medium} & \textbf{Slow} & \textbf{Average} \\
            \midrule
            \textbf{\textit{Baseline}}        & 2016 & 1749 & 3805 & 7222 & 3698 \\
            \textbf{\textit{+Fast123}}        & 348 (17.28\%) & 359 (20.51\%) & 842 (22.13\%) & 1568 (21.71\%) & 779 (21.07\%)
 \\
            \textbf{\textit{+Ours}}           & 204 (10.14\%)& 168 (9.60\%)& 448 (11.77\%) & 814 (11.27\%)& 409 (11.05\%)\\
            \bottomrule
        \end{tabular}
    \end{adjustbox}
\end{table}

\section{Conclusions}
\label{sec:5}
In this paper, we present a fast MMVD algorithm for VVenC based on filter difference analysis, which predicts MMVD gain using spatial gradients and residuals to avoid exhaustive search. Experimental results show better trade-off between encoding complexity and coding gains. Given that our method currently only supports the standard four cardinal directions; in future work, we plan to extend it to reduce the complexity of MMVD modes with more flexible diagonal angles \cite{Salehifar2022MMVD}.

\balance
\bibliographystyle{IEEEbib}
\bibliography{refs}

\end{document}